\documentclass{article}
\usepackage{spconf,amsmath,epsfig,amsfonts}
\usepackage{psfrag}
\usepackage{amsmath}
\usepackage{amssymb}
\usepackage{amsfonts}
\usepackage{color}
\usepackage{tabularx}

\newcolumntype{C}[1]{>{\centering\arraybackslash}p{#1}}

\def\punit{\, \mathrm}

\usepackage{subfigure}
\renewcommand{\thesubfigure}{\thefigure.\arabic{subfigure}}
\makeatletter
\renewcommand{\p@subfigure}{}
\renewcommand{\@thesubfigure}{{\bf Fig. \thesubfigure}.\ }
\makeatother

\title{Spatio-Temporal Prediction in Video Coding by Best Approximation}
\name{J\"urgen~Seiler, Haricharan~Lakshman, and Andr\'e~Kaup\vspace{-0.25cm}}
\address{Chair of Multimedia Communications and Signal Processing, \\University of Erlangen-Nuremberg,\\Cauerstr. 7, 91058 Erlangen, Germany\\ \{seiler, kaup\}@LNT.de, \  haricharan.lakshman@hhi.fraunhofer.de\vspace{-0.35cm}}

\begin{document}
\topmargin=0mm
\maketitle


\begin{abstract} \label{abstract}\vspace{-1.0mm}
Within the scope of this contribution we propose a novel efficient spatio-temporal prediction algorithm for video coding. The algorithm operates in two stages. First, motion compensation is performed on the block to be predicted in order to exploit temporal correlations. Afterwards, in order to exploit spatial correlations, this preliminary estimate is spatially refined by forming a joint model of the motion compensated block and spatially adjacent already decoded blocks. Compared to an earlier refinement algorithm, the novel one only needs very few iterations, leading to a speedup of factor 17. The implementation of this new algorithm into the H.264/AVC leads to a maximum reduction in data rate of up to nearly 13\% for the considered sequences. 
\end{abstract}




\section{Introduction} \label{sec:introduction} \vspace{-2.0mm}
In the past few years, the amount of video data  transmitted over digital channels has steadily increased. For this it is necessary that the video sequences are compressed by an encoder in order to reduce the data rate. Fortunately, video sequences can be strongly compressed. Most modern hybrid video codecs as e.\ g.\ the H.264/AVC \cite{Wiegand2003} use two different strategies for compressing the video sequence: prediction of the video signal to be coded and entropy coding of the prediction residual and the side information. Within the scope of this contribution we will focus on the first one. In this step, an estimate of the signal parts to be coded is generated from already transmitted areas, i.\ e.\ previous frames or already processed regions from the actual frame. Since only already transmitted areas are used for generating the estimate, the decoder can predict the signal in the same way as the encoder. Thus, instead of transmitting the quantized and entropy coded original video signal, only the quantized and entropy coded prediction error has to be transmitted. Therefore, besides the entropy coding, the abilities of a video codec directly depend on how efficiently the video signal to be coded can be predicted.

In most modern video codecs the prediction of the signal part to be coded is obtained by exploiting either temporal or spatial correlations. Thereby, spatial prediction is obtained by skillfully continuing the signal from already transmitted regions into the region being processed. On the other hand, the temporal prediction is performed by applying motion compensation (MC) on the region being coded, as described in \cite{Dufaux1995}. For this, a region in a previous frame is sought that fits the area to be coded best. The displacement of this region then is transmitted to the decoder as side information and the decoder then can form the prediction signal by taking the corresponding region from the already decoded frames. Although modern encoders can adaptively switch between spatial and temporal prediction in order to form the best predictor, a combined usage of temporal and spatial correlations only is applied rarely. Only few existent prediction algorithms exploit both correlations at the same time. As examples for this group of algorithms the \lq Inter Frame Coding with Template Matching Spatio-Temporal Prediction\rq\ by \cite{Sugimoto2004} or the \lq Pixelwise Adaptive Spatio-Temporal Prediction\rq\ by \cite{Day2008} can be mentioned.

In \cite{Seiler2008c} we proposed a new spatio-temporal prediction algorithm, the spatial refinement of motion compensation by Frequency Selective Approximation (FSA). This algorithm is able to reduce data rate significantly compared to pure temporal prediction. Unfortunately, this algorithm needs many iterations to form an adequate predictor and thus is computationally very expensive. We now want to propose a new algorithm for spatial refinement, the Relaxed Best Approximation (RBA), which is able to generate the model nearly as effectively as the original algorithm but with only very few iterations needed. The algorithm is based on the \lq Best Approximation\rq\ proposed by \cite{Kaup1998}.

In the next sections we will outline the idea of spatial refinement with FSA and especially with  the new algorithm in detail. We will also prove its abilities to improve the prediction quality in simulations with the H.264/AVC encoder and will show the reduction in computational complexity compared to the algorithm presented in \cite{Seiler2008c}.


\vspace{-1.5mm}\section{Problem Formulation and Background} \label{sec:probst}
\begin{figure}
	\psfrag{m}[t][t][0.8]{$m$}%
	\psfrag{n}[t][t][0.8]{$n$}%
	\psfrag{Agb}[l][l][0.8]{$\mathcal{A} = \mathcal{R} \cup \mathcal{B}$}%
	\psfrag{Block}[l][l][0.8]{$\mathcal{B}$}%
	\psfrag{Rgb}[l][l][0.8]{$\mathcal{R}$}%
	\psfrag{Pgb}[l][l][0.8]{$\mathcal{P}$}%
	\centering
	\includegraphics[width=0.2\textwidth]{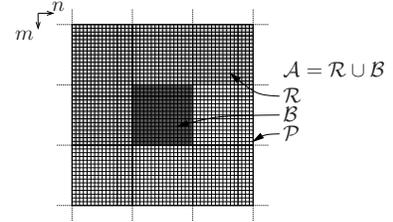}
	\emph{\caption{Projection area $\mathcal{P}$ containing the approximation area $\mathcal{A}$ consisting of area $\mathcal{R}$ subsuming the reconstructed blocks and the block $\mathcal{B}$ to be predicted}
	\label{fig:approx_area}}\vspace{-0.1cm}
\end{figure}

\vspace{-2.0mm}
We consider a block based video coder, operating in line scan order. Let the block actually being processed be denoted by area $\mathcal{B}$. This block is joined by $4$ blocks that have already been transmitted and are known to the decoder as well. These blocks are subsumed in area $\mathcal{R}$. We now regard the so called projection area $\mathcal{P}$, shown in Fig. \ref{fig:approx_area}, of $3\times 3$ blocks centered by the block $\mathcal{B}$. Besides $\mathcal{R}$ and $\mathcal{B}$, this square area contains four blocks that have not been coded yet. The novel idea of the spatial refinement proposed in \cite{Seiler2008c} is that first a preliminary temporal extrapolation is formed by motion compensation for the block $\mathcal{B}$. By transmitting the motion vector as side information, the decoder can perform the motion compensation in the same way. In a second step, a model is generated for the union $\mathcal{A}=\mathcal{R}\cup\mathcal{B}$, called approximation area. Finally the samples corresponding to $\mathcal{B}$ are taken from the model and are used as predictor. As the model incorporates information from the temporally extrapolated block $\mathcal{B}$ as well as from the spatially adjacent blocks $\mathcal{R}$ this will form a better predictor than the purely temporal one.

Let the intensities of the samples in area $\mathcal{P}$ be denoted by $f\left[m, n\right]$ and the model, representing the refined signal, be denoted by $g\left[m, n\right]$. $\left(m,n\right)$ represent the spatial coordinates and area $\mathcal{P}$ is of size $M\times N$ samples. The parametric model 
\begin{equation}
g\left[m,n\right]=\sum_{k\in\mathcal{K}}c_k\varphi_k\left[m,n\right]
\end{equation}
emanates from a weighted superposition of the two-dimen\-sional basis functions $\varphi_k\left[m,n\right]$ with appropriate weights $c_k$. The set $\mathcal{K}$ covers all basis functions used for modeling.

For the purpose of spatial refinement of the motion compensated signal, the samples of $\mathcal{R}$ that are close to $\mathcal{B}$ need to have more influence than the ones far away. This non-uniform influence is incorporated by means of the later used weighting function 
\begin{equation}
w\left[m,n\right] = \left\{\begin{array}{ll} \mu & , \forall \left(m,n\right) \in \mathcal{B} \\ \hat{\rho}^{\sqrt{\left(m-\frac{M-1}{2}\right)^2 + \left(n-\frac{N-1}{2}\right)^2}} & , \forall \left(m,n\right) \in \mathcal{R} \\0 & , \mbox{ else } \end{array} \right.
\end{equation}
Hence the block $\mathcal{B}$ gets the constant weight $\mu$ and the samples in $\mathcal{R}$ get an exponentially decreasing weight with an increasing distance, controlled by the decay factor $\hat{\rho}$. 

The model $g\left[m,n\right]$ now should be generated in such a way as to minimize the weighted approximation error energy 
\begin{equation}
E=\sum_{\left(m,n\right) \in \mathcal{P}}w\left[m,n\right](f\left[m,n\right]-g\left[m,n\right])^2.
\label{werren}
\end{equation}
According to (\ref{werren}), the model generation for spatio-temporal prediction can be viewed as an error minimization task. It is important to notice that the traditional approach of minimizing the error by taking partial derivatives with respect to the unknown coefficients $c_k$ and equating them to zero leads to an underdetermined system of equations because the number of known samples is less than the total number of points considered. For such problems, as shown in \cite{Olshausen1997}, sparsity based solutions are advocated because they are capable of capturing important characteristics of a signal. However, direct solutions using $l_0$ quasi-norm as a sparsity measure are NP-Hard according to \cite{Natarajan1995}.

The Frequency Selective Approximation (FSA) algorithm from \cite{Kaup1998} employed in \cite{Seiler2008c} is an iterative error minimization procedure which produces a sparse solution. It belongs to the class of Greedy Approximation techniques in which the signal is approximated in terms of one additional basis function per iteration. This involves the selection of a basis function $\varphi_u\left[m,n\right]$ and the computation of the optimal expansion coefficient $c_u$ corresponding to the selected basis function. In each iteration, the basis function is chosen in such a way that the reduction of the weighted residual energy is maximized. After generating the parametric model $g\left[m,n\right]$, the area of interest is cut out and used as predictor for the block being coded. For a detailed discussion of FSA, please refer to \cite{Seiler2008c, Kaup1998}.

\begin{figure}
	\begin{center}
		\includegraphics[width=0.45\textwidth]{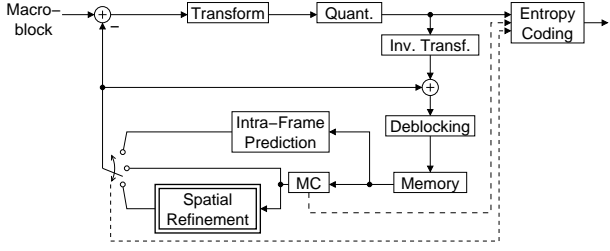}
	\end{center}\vspace{-0.5cm}
	\emph{\caption{Block diagram of a hybrid video encoder with spatial refinement}\label{fig:bd}}\vspace{-0.1cm}
\end{figure}

The block diagram in Fig. \ref{fig:bd} shows the position of the spatial refinement step in a generalized hybrid video encoder. The deblocking filter, as e.\ g.\ used in the H.264/AVC \cite{Wiegand2003}, is an optional feature that can be used in addition to the spatial refinement without interfering with it. As shown in \cite{Seiler2008c}, the gain obtainable by deblocking adds to the gain obtainable by spatial refinement. This is since deblocking only improves the reference frames and therewith motion compensation but does not incorporate spatial correlations for prediction.


\vspace{-1.5mm}\section{Best Approximation} \label{sec:approximation}\vspace{-2.0mm}
This section introduces the Best Approximation (BA), originally proposed in \cite{Kaup1998} and discusses its advantages. The basic idea of BA is close to FSA as in every iteration step, one basis function is added to the model generated so far. The basis function selected is, as in FSA, the one that maximizes the approximation error energy decrement. But unlike FSA where the residuum is approximated just by the selected basis function in each iteration step, BA modifies the expansion coefficients of all the already selected basis functions in order to produce the best possible approximation using the selected set. The expansion coefficients for the selected basis functions are calculated by solving a projection problem in least squares sense. According to approximation theory, such an algorithm can be categorized as an Orthogonal Greedy Approximation whose anatomy is similar to Greedy Approximation but has a faster convergence \cite{Temlyakov2000}. 

Let $g^{\left(\nu\right)}\left[m,n\right]$ represent the parametric model and $\mathcal{K}^{\left(\nu\right)}$ the set of selected basis functions in the $\nu$-th  iteration step. The new model, in which the coefficients of all the selected basis functions are updated, can be written as
\begin{equation}
g^{(\nu+1)}\left[m,n\right]=g^{\left(\nu\right)}\left[m,n\right] + \hspace{-3mm} \sum_{u \in \mathcal{ K}^{\left(\nu+1\right)}} \hspace{-2.5mm} \Delta c_u \varphi_u\left[m,n\right].
\label{eq:model}
\end{equation}
The weights $\Delta c_u$ can be computed by setting the partial derivatives of the weighted error energy with respect to all $\Delta c_u$ to zero. This yields a system of linear equations of size $\left| \mathcal{K}^{\left(\nu+1\right)}\right|$ for the coefficients $\Delta c_u$
\[
\hspace{-15mm}\sum_{\left(m,n\right) \in \mathcal{P}} \hspace{-2.5mm} w\left[m,n\right](f\left[m,n\right]-g^{\left(\nu\right)}\left[m,n\right]) \varphi_k\left[m,n\right] =
\]
\vspace{-2.5mm}
\begin{equation} 
\sum_{u \in \mathcal{K}^{\left(\nu+1\right)}} \hspace{-4mm}\Delta c_u \hspace{-2mm} \sum_{\left(m,n\right) \in \mathcal{P}} \hspace{-2mm} w\left[m,n\right] \varphi_k\left[m,n\right]\varphi_u\left[m,n\right] , \ \ \forall k \in \mathcal{K}^{\left(\nu+1\right)}
\label{eq:bbasys}
\end{equation}
Solving this linear system gives all the coefficients $\Delta c_u$ which are then used to update the parametric model
\begin{equation}
c_u^{\left(\nu+1\right)}=c_u^{\left(\nu\right)}+\Delta c_u, \  \forall \hspace{2mm} u \in {\cal K}^{\left(\nu+1\right)}.
\end{equation}
These steps of selecting one basis function and updating the expansion coefficients of all selected basis functions are repeated until a predefined maximum number of iterations is reached. By updating the expansion coefficients of all the selected basis functions in one iteration step, the algorithm requires a smaller number of iterations compared to FSA. 

\vspace{-1.5mm}\subsubsection*{Relaxation Scheme}\vspace{-1.5mm}
The spatial refinement of temporally predicted data depends on the ability of the parametric modeling to combine the important characteristics of the regions $\mathcal{R}$ and $\mathcal{B}$. Although BA yields the best possible approximation of union $\mathcal{R}\cup\mathcal{B}$ in each iteration step, it might not result in a better spatial refinement. Additionally, in each iteration of BA, a system of linear equations needs to be solved, which adds computational complexity. In order to tackle these issues, we introduce a relaxation scheme which does not only improve the refinement performance of BA but also provides a reduction in complexity. The new scheme performs a Relaxed Best Approximation (RBA) by selecting all the basis functions that provide at least a specified fraction of the maximum reduction in error energy into the chosen set for a particular iteration. Therefore, the relaxation parameter $\tau$ between $0$ and $1$ is introduced that controls which basis functions to be added to the model in a certain iteration step. Thus, the steps of RBA are defined as follows:

\begin{enumerate}
\item Select all basis functions $\mathbf{\varphi}_u\left[m,n\right]$ that satisfy
\begin{equation}
\Delta E_{\mathbf{\varphi}_u}^{\left(\nu\right)} \geq \tau \cdot \max_{\mathbf{\varphi}_k} \Delta E_{\mathbf{\varphi}_k}^{\left(\nu\right)}
\end{equation}
with $\Delta E_{\mathbf{\varphi}_u}^{\left(\nu\right)}$ being the reduction in residual energy by selecting basis function $\varphi_u$
\item Update model $g^{\left(\nu+1\right)}\left[m,n\right]$ according to (\ref{eq:bbasys}) and (\ref{eq:model})
\item Compute new residual and iterate
\end{enumerate} 

So, by using RBA, in every iteration step, several basis functions can be added to the model. This results in a reduction of the number of iterations needed to set up the model for forming the predictor and with that the overall complexity will become a lot smaller than with FSA or BA.


\vspace{-1.5mm}\section{Simulation Setup and Results} \label{sec:results}\vspace{-2.0mm}
In order to evaluate the abilities of the proposed algorithm, we implemented this new prediction mode into the H.264/AVC reference software JM 10.2, Baseline Profile, Level 2.0~. For motion compensation, quarter pixel accuracy is applied at a search range of $16$ pixels and one reference frame. For comparing the refined prediction with the original pure motion compensation and spatial refinement by FSA, the rate control is switched off and $10$ fixed QPs from $16$ to $43$ are used.

In order to evaluate the prediction quality, the sequences ``Crew'', ``Foreman'' and ``Vimto'' in CIF are encoded at $30$ frames per second with three different settings: pure motion compensation for prediction, spatial refinement by FSA \cite{Seiler2008c} and spatial refinement by the proposed RBA. As the spatial refinement might not increase the prediction quality for every macroblock, the encoder has to compare the refined and the unrefined prediction signal with the original block and has to signal the decoder if the refinement step is applied. To account for this, one bit per macroblock is added to the data for the two refinement algorithms as a worst case assessment for the emerging additional side information.

The weighting function $w\left[m,n\right]$ used for spatial refinement is the same for FSA and RBA with the decay factor $\hat{\rho}$ chosen to $0.8$ and the weighting of the motion compensated block chosen to $\mu=0.5$. According to \cite{Seiler2008c}, for both the algorithms, the set of basis functions used for model generation are the functions of the two-dimensional Fourier transform, since this set is especially suited for natural images. According to the previous experiments, FSA uses $200$ iterations, whereas the proposed RBA uses only $4$ iterations to form the model. For RBA the factor $\tau$ is set to $0.5$ and $20$ basis functions are maximally added to the model in one iteration step. Fortunately, the above mentioned parameters all are not very critical and can be varied in a relatively wide range without heavily affecting the prediction performance.

\begin{figure}
	\psfrag{s05}[t][t][0.9]{\color[rgb]{0,0,0}\setlength{\tabcolsep}{0pt}\begin{tabular}{c}Rate $\left[\punit{kbit}/\punit{s}\right]$\end{tabular}}%
	\psfrag{s06}[b][b][0.9]{\color[rgb]{0,0,0}\setlength{\tabcolsep}{0pt}\begin{tabular}{c}$\punit{PSNR} \left[\punit{dB}\right]$\end{tabular}}%
	\psfrag{s08}[b][b][0.9]{\color[rgb]{0,0,0}\setlength{\tabcolsep}{0pt}\begin{tabular}{c}\end{tabular}}%
	\psfrag{s10}[][][0.9]{\color[rgb]{0,0,0}\setlength{\tabcolsep}{0pt}\begin{tabular}{c} \end{tabular}}%
	\psfrag{s11}[][][0.9]{\color[rgb]{0,0,0}\setlength{\tabcolsep}{0pt}\begin{tabular}{c} \end{tabular}}%
	\psfrag{s12}[l][l][0.75]{\color[rgb]{0,0,0}Vimto, 2D-RBA}%
	\psfrag{s13}[l][l][0.75]{\color[rgb]{0,0,0}Crew, MC}%
	\psfrag{s14}[l][l][0.75]{\color[rgb]{0,0,0}Crew, 2D-FSA}%
	\psfrag{s15}[l][l][0.75]{\color[rgb]{0,0,0}Crew, 2D-RBA}%
	\psfrag{s16}[l][l][0.75]{\color[rgb]{0,0,0}Foreman, MC}%
	\psfrag{s17}[l][l][0.75]{\color[rgb]{0,0,0}Foreman, 2D-FSA}%
	\psfrag{s18}[l][l][0.75]{\color[rgb]{0,0,0}Foreman, 2D-RBA}%
	\psfrag{s19}[l][l][0.75]{\color[rgb]{0,0,0}Vimto, MC}%
	\psfrag{s20}[l][l][0.75]{\color[rgb]{0,0,0}Vimto, 2D-FSA}%
	\psfrag{s21}[l][l][0.75]{\color[rgb]{0,0,0}Vimto, 2D-RBA}%
	\psfrag{x12}[t][t][0.9]{$0$}%
	\psfrag{x13}[t][t][0.9]{$500$}%
	\psfrag{x14}[t][t][0.9]{$1000$}%
	\psfrag{x15}[t][t][0.9]{$1500$}%
	\psfrag{x16}[t][t][0.9]{$2000$}%
	\psfrag{x17}[t][t][0.9]{$2500$}%
	\psfrag{x18}[t][t][0.9]{$3000$}%
	\psfrag{v12}[r][r][0.9]{$30$}%
	\psfrag{v13}[r][r][0.9]{$35$}%
	\psfrag{v14}[r][r][0.9]{$40$}%
	\psfrag{v15}[r][r][0.9]{$45$}%

	\centering 
	\vspace{-0.75cm}\includegraphics[width=0.42\textwidth]{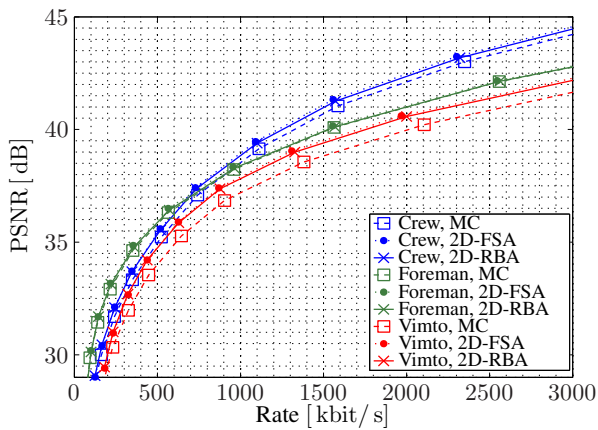}
	\emph{\caption{RD-curves for the first $99$ P-frames of the used CIF-sequences at $30$ frames per second with direct motion compensation (MC) and spatial refinement by FSA \cite{Seiler2008c} and RBA}\label{fig:rd_curve}}\vspace{-0.1cm}
 \end{figure}

\begin{table}
\footnotesize 
\centering
\begin{tabular}{|c|c|c|c|c|}
\hline
Sequence & \multicolumn{2}{c|}{FSA \cite{Seiler2008c}} & \multicolumn{2}{c|}{RBA}  \\ \hline
& Avg. Rate & Avg. $\punit{PSNR}$ &  Avg. Rate & Avg. $\punit{PSNR}$ \\ 
& Reduction & Gain &  Reduction & Gain\\ \hline
``Crew'' & $7.31\%$ & $0.37\punit{dB}$ & $6.20\%$ & $0.31\punit{dB}$ \\ \hline
\hspace{-1mm}``Foreman''\hspace{-1mm} & $3.20\%$ & $0.13\punit{dB}$ & $1.42\%$ & $0.06\punit{dB}$ \\ \hline
``Vimto'' & $13.42\%$ & $0.66\punit{dB}$ & $12.61\%$ & $0.62\punit{dB}$ \\ \hline
\end{tabular}
\emph{\caption{Achievable average relative rate reduction and average $\punit{PSNR}$ gain according to \cite{Bjontegaard2001} for spatial refinement by FSA  and the proposed RBA}\label{tab:psnr_gain}}\vspace{-0.1cm}
\end{table}

In Fig. \ref{fig:rd_curve} the rate-distortion curves for the first $99$ P-frames of the considered sequences are shown. For each sequence the figure contains the curves for the cases that only pure motion compensation is applied for prediction and that spatial refinement is used, either with FSA or the proposed RBA. Obviously, both refinement algorithms lead to a reduction in data rate needed to obtain a certain quality. Unfortunately, for the sequence ``Foreman'' this gain is small and cannot be seen well in the figure. Hence and to quantify the gain, Tab. \ref{tab:psnr_gain} lists the average rate reduction and average $\punit{PSNR}$ gain compared to motion compensation. Both averages are calculated according to \cite{Bjontegaard2001} and one can see a maximum reduction in data rate of up to $13\%$ and a mean reduction of about $7\%$ for the regarded sequences.

Comparing the improvement introduced by the spatial refinement with FSA and RBA, it becomes apparent, that FSA is slightly better than RBA. But one major drawback of FSA is the large number of iterations needed to generate the model. For this reason Tab. \ref{tab:calctime_reduction} shows the mean calculation time per frame for the spatial refinement for both algorithms. The spatial refinement step was carried out in MATLAB v7.6 on a Intel Core 2 @ $2.4\punit{GHz}$. The motion compensated block and the spatially adjacent blocks are exported to MATLAB for refinement. Afterwards the refined block is retransfered to JM for the further coding steps. According to Tab. \ref{tab:calctime_reduction}, we can see that RBA needs only about $1/17$-th of the processing time of FSA, which could be further improved by solving the system of equations from (\ref{eq:bbasys}) more efficiently. Considering this large reduction in processing time, the small increment in data rate compared to FSA becomes negligible.

\begin{table}
\footnotesize 
\centering
\vspace{-0.25cm}\begin{tabular}{|c|c|c||c|}
\hline
Sequence & \multicolumn{2}{c||}{Mean time / frame } & Time Gain \\ \hline
& FSA \cite{Seiler2008c} & RBA &  \\ \hline
``Crew''& $ 217.49\punit{sec}$ & $13.03\punit{sec}$ & $ 16.69 $ \\ \hline
``Foreman''& $214.93\punit{sec}$ & $12.15\punit{sec}$ & $17.69$ \\ \hline
``Vimto''& $217.07 \punit{sec}$ & $12.43\punit{sec}$ & $17.46$ \\ \hline
\end{tabular}
\emph{\caption{Mean processing time per frame for spatial refinement}\label{tab:calctime_reduction}}\vspace{-0.1cm}
\end{table}


\vspace{-1.5mm}\section{Conclusion} \label{sec:conclusion}\vspace{-2.0mm}
Regarding the above presented results, one can easily see that prediction in video coding can significantly gain from using spatial as well as temporal information to form the predictor. Within the scope of this contribution we presented a novel spatial refinement algorithm which produces an average reduction of the data rate of up to $13 \%$ at maximum for the regarded sequences. In addition, this new algorithm needs only very few iterations to form the predictor and is about $17$ times faster than an earlier proposed algorithm. 

Furthermore, our current research is focused on combining the spatial refinement with more sophisticated prediction techniques and on a further complexity reduction.





{\ninept\renewcommand{\baselinestretch}{0.8}

}

\end{document}